\documentclass{sf2a-conf2011}
\usepackage{graphicx}
\usepackage{hyperref}
\usepackage[]{natbib}  
\usepackage[cyr]{aeguill}
\usepackage{epstopdf}

\def\BibTeX{{\rm B\kern-.05em{\sc i\kern-.025em b}\kern-.08em
    T\kern-.1667em\lower.7ex\hbox{E}\kern-.125emX}}
\bibpunct{(}{)}{;}{a}{}{,}  %%%%%%%%%%%%%  A&A bibliography style
\begin{document}
\TitreGlobal{SF2A 2011}
%%-----------------------------------------------------------------
%%      the top matter
%%
\title{Some recent results of the Codalema Experiment}
\runningtitle{Recent results of the Codalema Experiment}
\author{Rebai, A.}\address{SUBATECH IN2P3-CNRS/Universit\'e de Nantes/\'Ecole des Mines de Nantes France}
\author{the CODALEMA collaboration}\address{LESIA, Observatoire de Paris-Meudon France - Station de Radioastronomie de Nan\c cay France}
%% Keep this line, even if the page will be settled afterwards.
\setcounter{page}{237}
%%-----------------------------------------------------------------
\maketitle
%%-----------------------------------------------------------------
%%        The abstract
%% 
%%  Warning!  within the abstract:
%%  - do not use macros. 
%%  - do not use commands like: \cite, \citet, \citep ... etc.

\begin{abstract}
Codalema is one of the experiments devoted to the detection of ultra high energy cosmic rays by the radio method. The main objective is to study the features of the radio signal induced by the development in the atmosphere of extensive air showers (EAS) generated by cosmic rays in the energy range of $10^{16}$-$10^{18}~eV$. After a brief presentation of the detector features, the main results obtained are reported (emission mechanism, lateral distribution of the electric field, energy calibration, etc.). The first studies of the radio wave front curvature are discussed as new preliminary results.
\end{abstract}

%% Insert the keywords (to appear in the ADS indexing)
%% Keywords must be separated by a comma


%%-----------------------------------------------------------------
\section{Introduction}
%%---------------------
A century after the discovery of cosmic rays, several fundamental issues related to the nature and the origin of ultra high energy cosmic rays (UHECR) remain unanswered, despite intensive experimental efforts. The main difficulty opposing the progress is due to the extremely low flux of UHECR (1 particle/$km^2$/century at $10^{20}~eV$) and the present performances of particle detectors arrays and fluorescence telescopes \citep{AugerComp2010PRL,HiresComp2010PRL}. In the recent decade, the measurement of the radio counterparty of EAS becomes a promising technique. Many scientific collaborations like CODALEMA in France \citep{GeoMagOrigCodAstroPh} and LOPES in Germany \citep{LDFRadioSigEasLOPESAstroPh} have demonstrated the feasibility of this method to deduce the EAS features. The potentialities of the radiodetection resides in several avantages, among them may be mentioned: the operating duty cycle close to $100\%$, the sensitivity to the shower longitudinal development in the atmosphere and the mechanical robustness, the simplicity and the low cost of the antennas. The CODALEMA experiment is installed at the radio observatory site (Nan\c cay, 47.3$^{\circ}$N, 2.1$^{\circ}$E and 137 m above sea level). Its main goal is to improve the pioneer experimental results \citep{AllanRevue}, taking avantage of ultrafast electronic devices and a quiet radio environment from anthropic transmitters in the detection bandwidth \citep{RadioElecFeaturEASCodaAstroPh}. From the phenomenological point of view, CODALEMA has made progress in the understanding of radio signal origin, showing that the geomagnetic field is the main actor in radio signal emission via the geomagnetic mechanism \citep{GeoMagOrigCodAstroPh}, and showing recently the contribution of a second emission mechanism due to the shower negative charge excess \citep{ICRC2011MarinExcess}. This contribution reports on the last results of the experiment CODALEMA with updated data set, recalling the observations of the north-south asymmetry and of the energy correlation. Finally the reconstruction method of the radio wavefront radius of curvature is presented.
%%-----------------------
\section{Experimental situation}
The CODALEMA experiment is made of two main arrays of detectors. The first array is built with $24$ short active dipoles antennas distributed on a cross geometry with dimensions $400~m$ by $600~m$. This apparatus is used to study the EAS radio counterpart. The dipole antenna is made by two radiator arms each $60~cm$ long at a height of $1.2~m$. The antenna design was optimized to reach an almost isotropic pattern. Low noise amplifier (LNA) is used to amplify the electric signal. It is conceived to be sensitive to the radio galactic background and is linear over a wide frequency band from $0.1$ MHz up to $230$ MHz. The second apparutus is a ground-based particle detector array formed by $17$ plastic scintillators placed on a square with $340~m$ side. It measures the primary particle energy and provides the trigger signal to the other detector arrays. The entire acquisition system (DAQ) is trigged by the passage of secondary particles in coincidence through each of the five central scintillators.
\section{A North-South asymmetry: a geomagnetic effect signature in the production of the radio signal}
\begin{figure}[!h]
 \centering
\includegraphics[width=0.4\textwidth,clip]{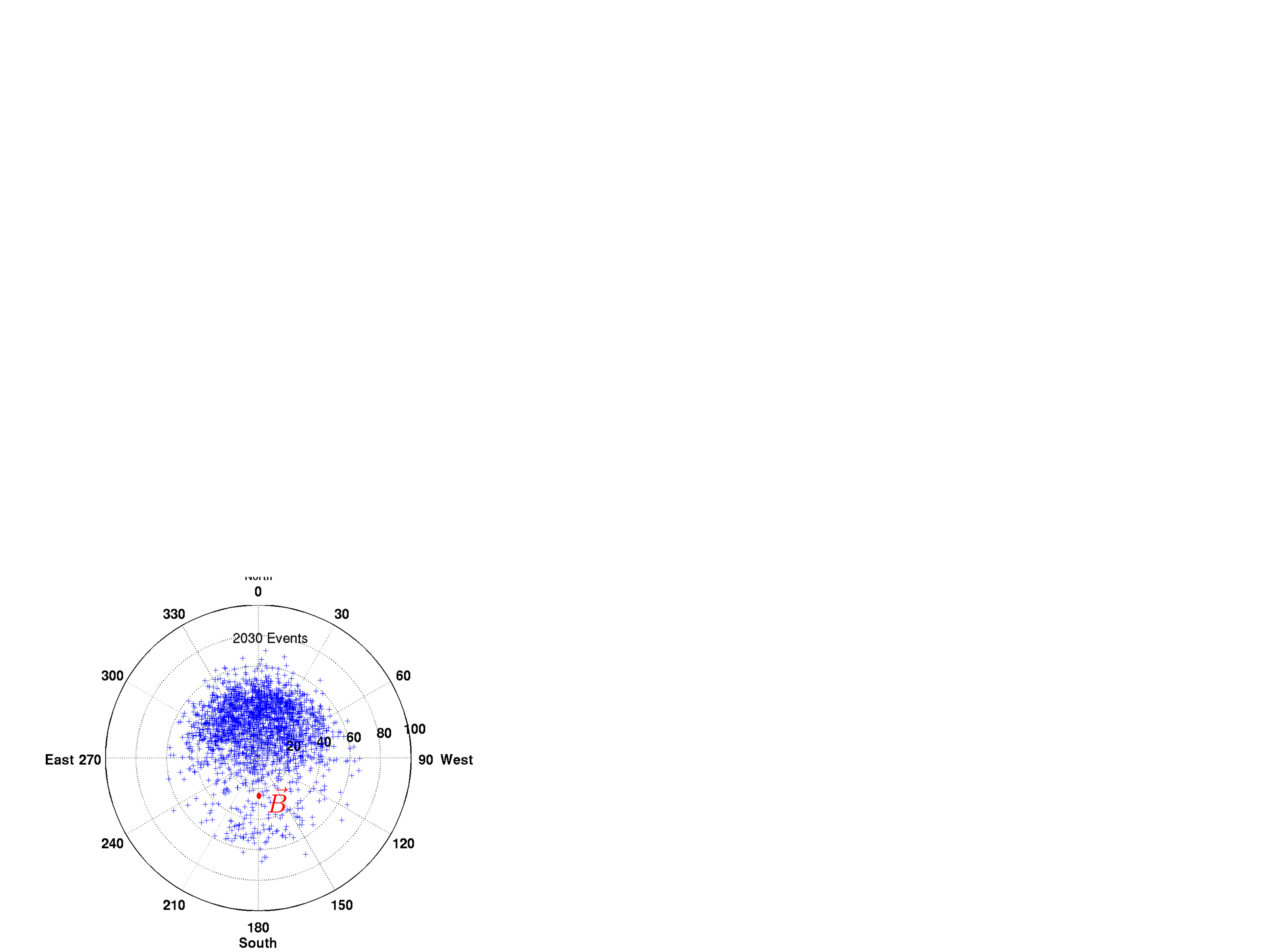}
\includegraphics[width=0.3\textwidth,clip]{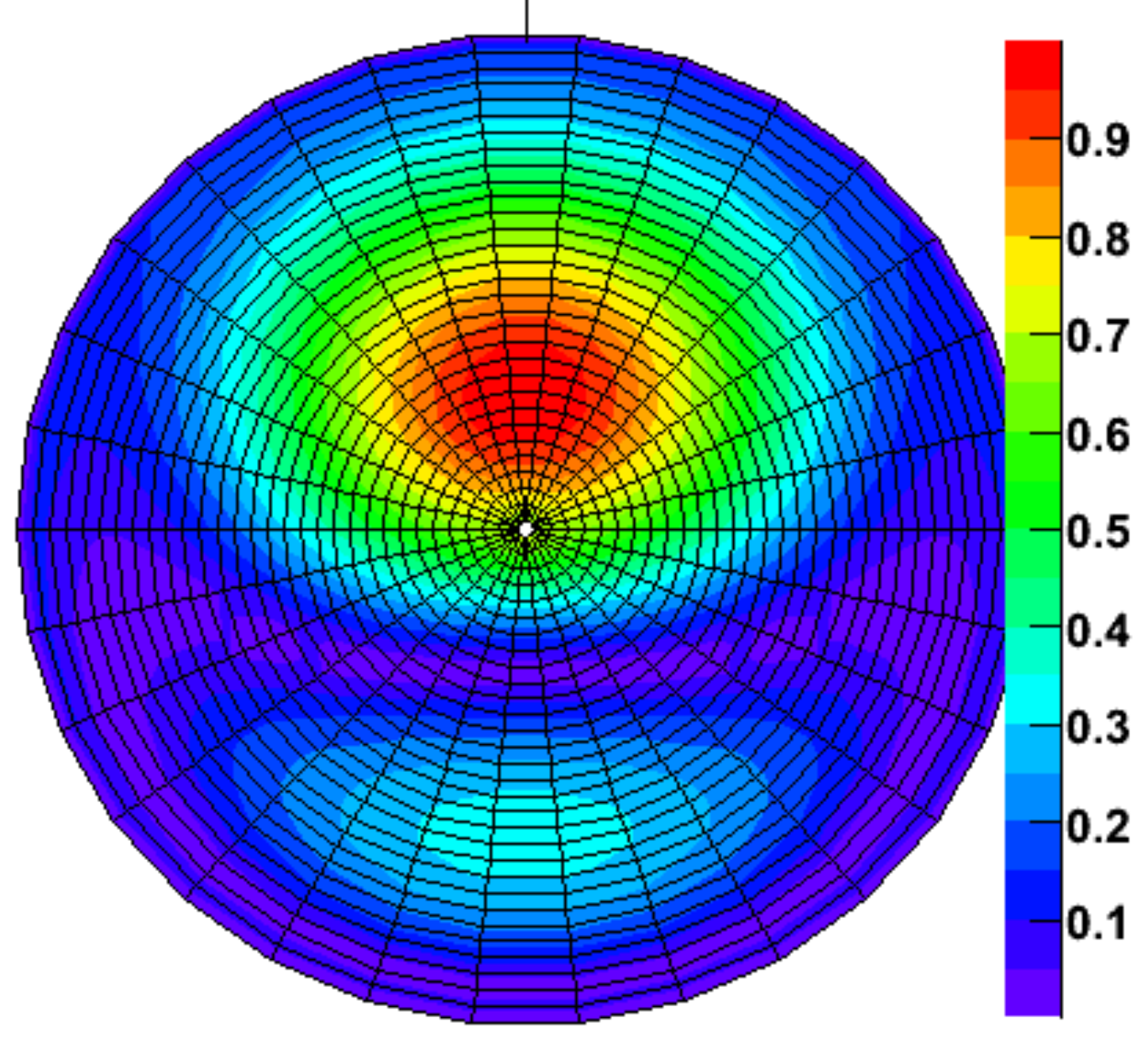}
\caption{{\bf Left:} A plot sky showed the arrival directions distribution of detected EAS events. The red dot indicates the local geomagnetic field ($\theta=27^\circ$,$\phi=0^\circ$). {\bf Right:} Sky map of the geomagnetic model theoretical prediction based on the projection of the product $\vec{v} \wedge \vec{B}$ on the East-West direction multiplied by the trigger coverage map. The colormap is normalized to 1.}
\label{NS_anisotropy}
\end{figure}

%%-------------------------
CODALEMA is the first experience which has reported a large and stable assymetry in counting rates between showers coming from north and south \citep{GeoMagOrigCodAstroPh,OlivierArena2010}. This asymmetry has been interpreted as a signature of the geomagnetic field effect in the radio signal emission process. The local geomagnetic field $\vec{B}$ is the main cause of this assymmetry through the action of Lorentz force on the secondary charged particles via the $\vec{v} \wedge \vec{B}$ term. We find that among the $2030$ events detected in coincidences with the two arrays: $1708$ events coming from the north and $322$ coming from the south (see Fig.~\ref{NS_anisotropy}, Left). The respective ratio equal to $0.188$ is consistent with the value of $0.17$ obtained with a previous statistic used in \citep{GeoMagOrigCodAstroPh}. A pure statistical effect is excluded with $15~\sigma$. One can enhance here on the fact that this asymmetry is stable over time and with different data samples.
%%-------------------------
\section{Energy calibration of the Codalema antenna array}
\begin{figure}[!h]
 \centering
\includegraphics[width=0.45\textwidth,clip]{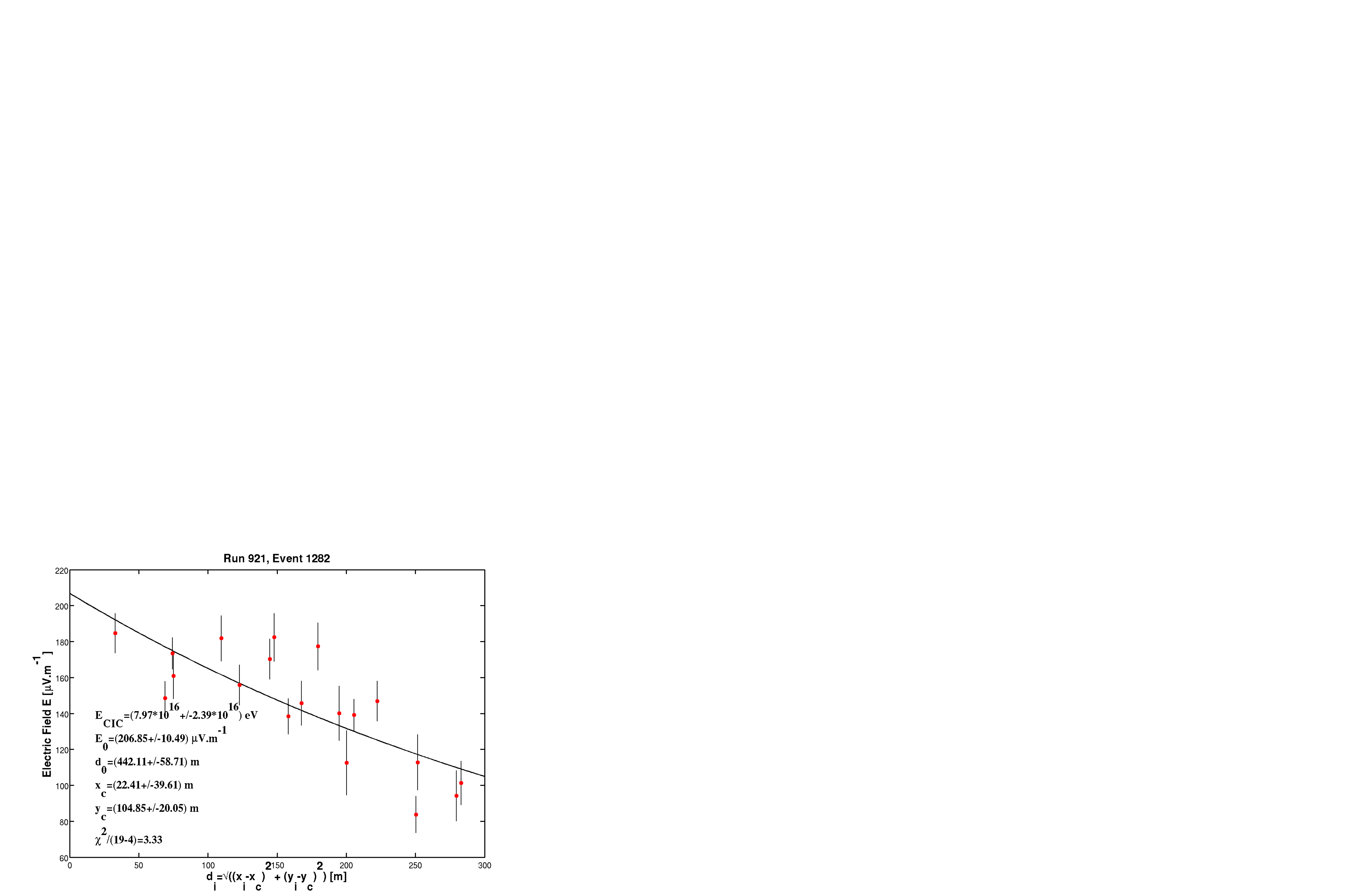}
\includegraphics[width=0.45\textwidth,clip]{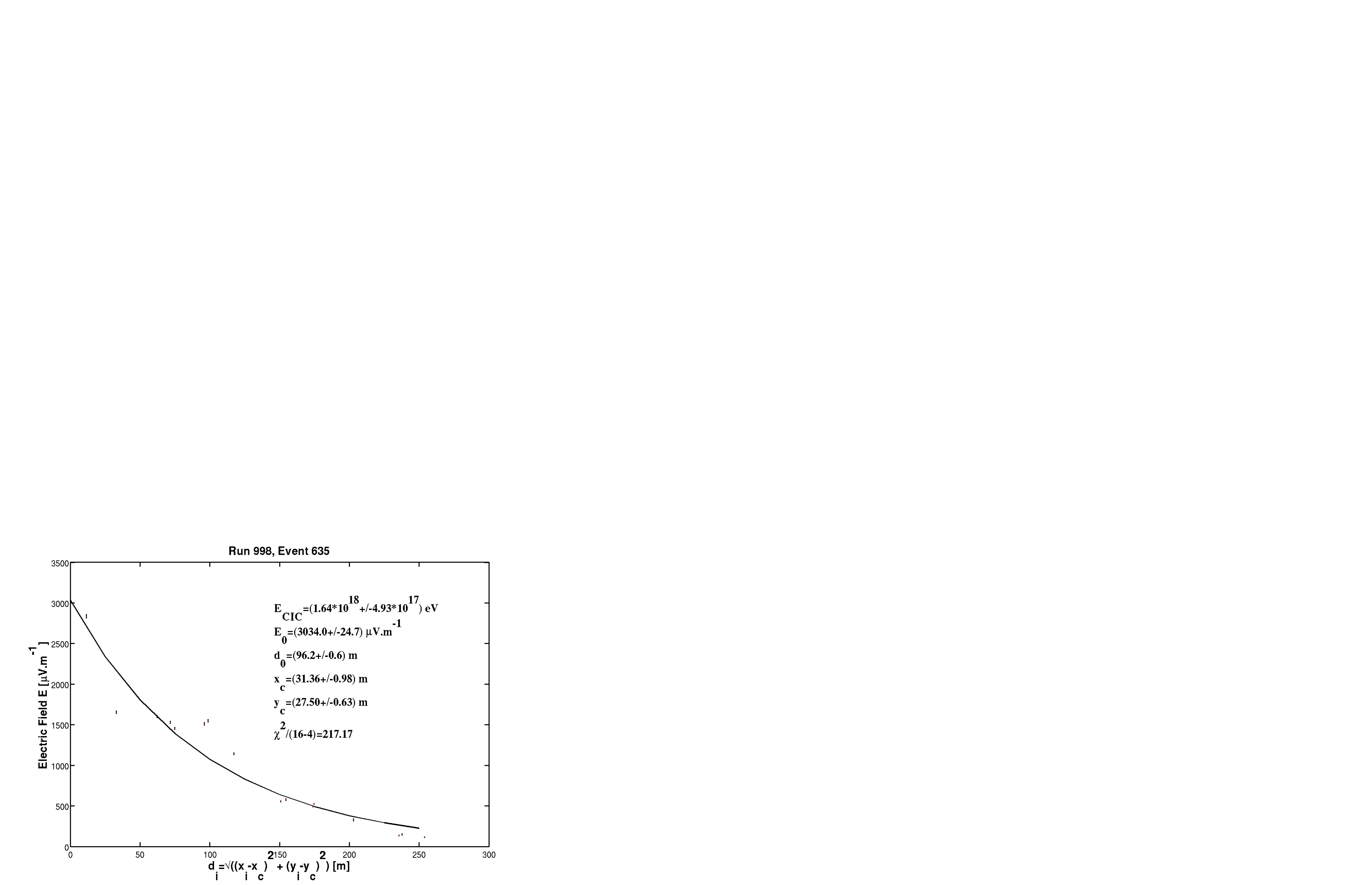}
\caption{Radio lateral distributions reconstructed from antenna signals. The full line show the result of an exponential law fit (Allan formula). {\bf Left:} We show the lateral distribution of an event detected by 19 antennas. The primary particle energy is estimated to $E_{p}=7.97~10^{16}~eV$. Error bars correspond to the radio galactic background. In this example, the signal amplitudes are close to the noise level. {\bf Right:} We show the lateral distribution of an other event ,detected by 16 antennas, with larger energy $E_p=1.64~10^{18}~eV$. In this case, error bars are small compared to signals.}
\label{LDFstudies}
\end{figure}

\begin{figure}[!ht]
\centering
\includegraphics[width=0.45\textwidth,clip]{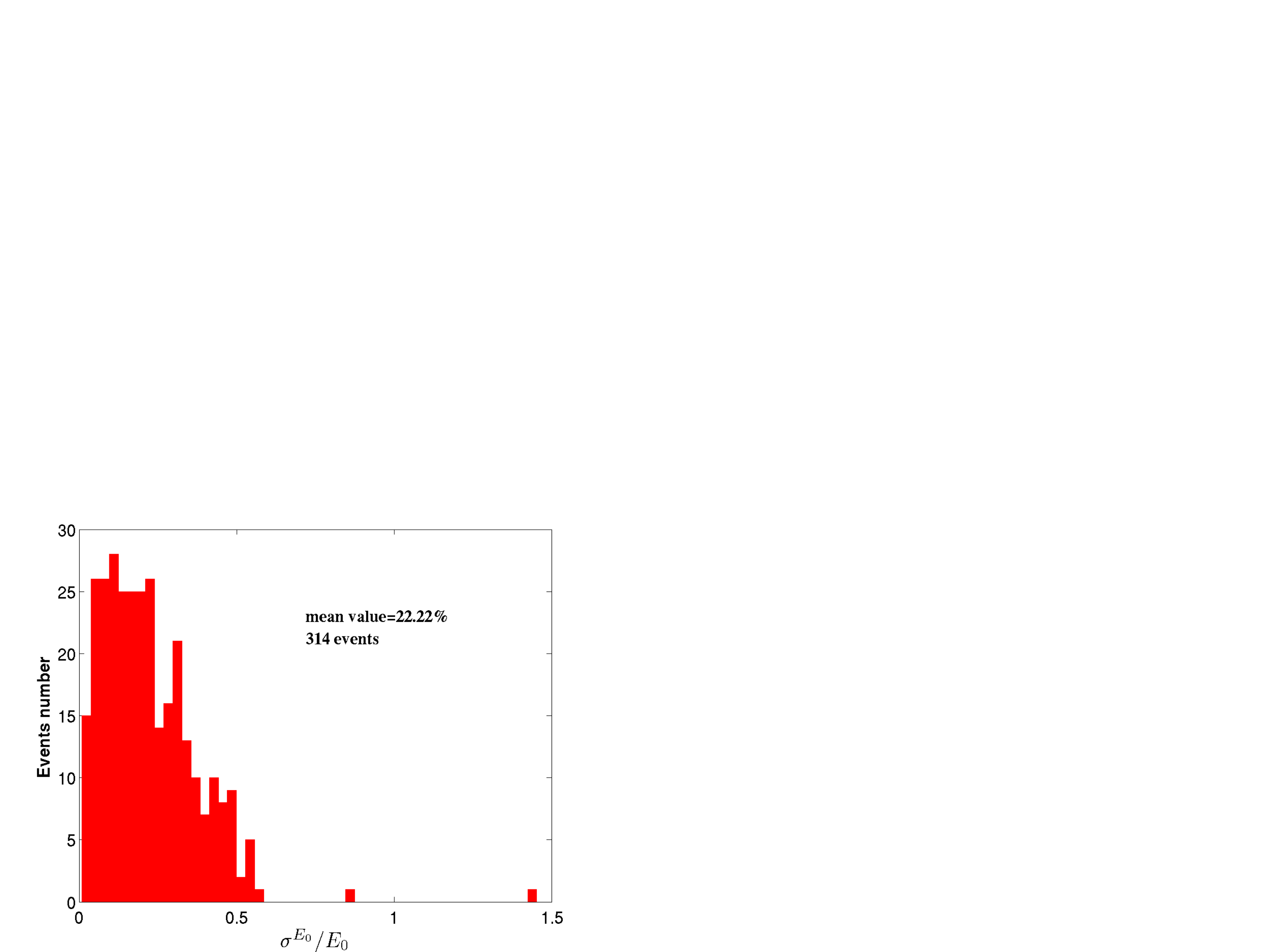}
\includegraphics[width=0.45\textwidth,clip]{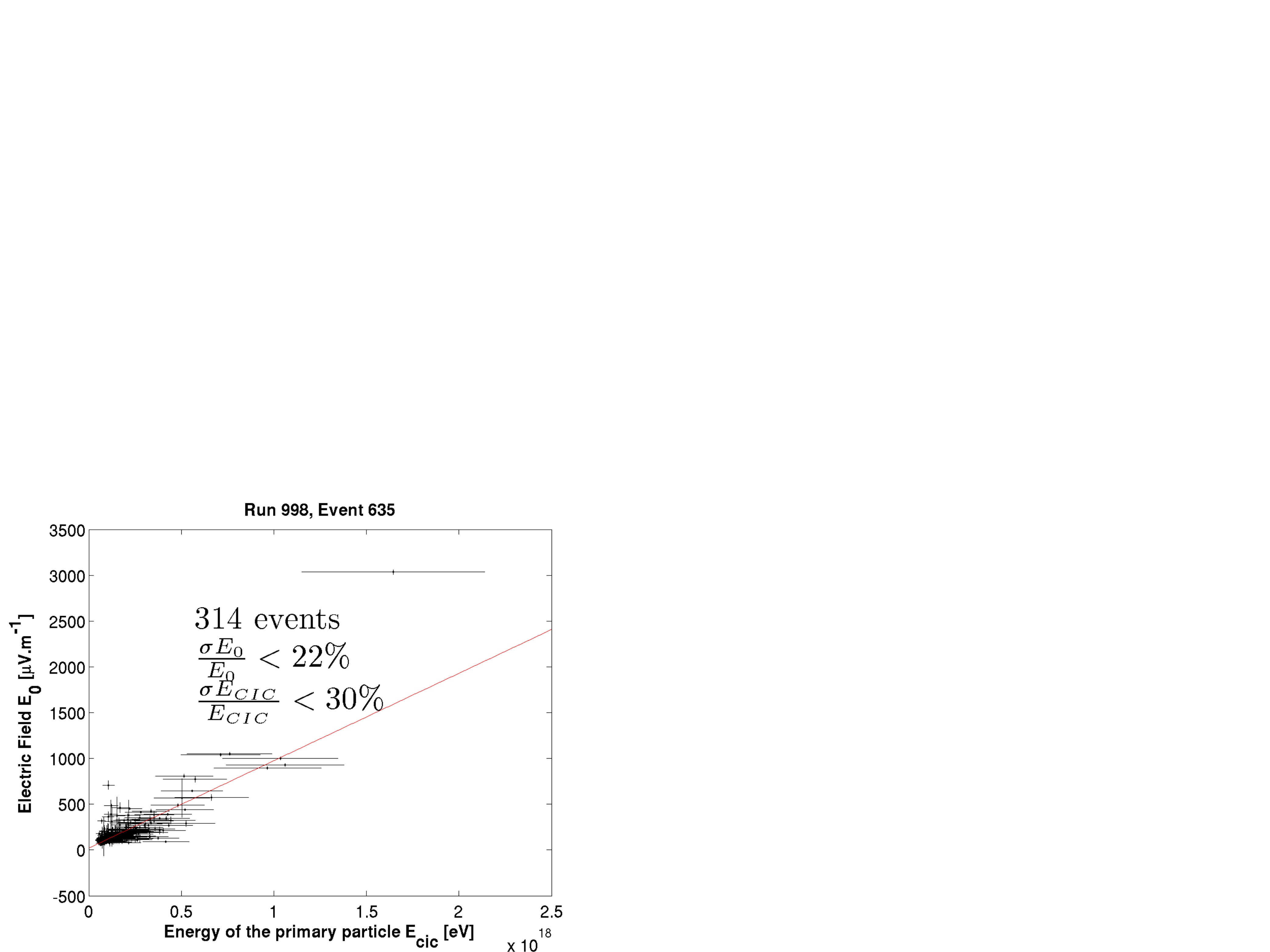}
\caption{{\bf Left:} $E_0$ statistical errors $\frac{\sigma E_0}{E_0}$ histogram of the LDF fit method. {\bf Right:} Fit of the correlation between the primary particle energy $E_p$ and the electric field at the shower axis $E_0$. This study strongly depends on both $E_p$ and $E_0$ errors.}
\label{Correlationstudies}
\end{figure}

%%-------------------------
The energy of the primary particle $E_p$ is one of the most important parameters for studying EAS. To avoid the use of the particles detectors array, the radio technique must demonstrate its ability to estimate the primary particle energy with the information given by the radio signal alone. In this perspective, a study of the correlation between the shower energy and the electric field created at the shower axis $E_{0}$ is the natural way to determine the energy calibration response of the antenna array. In this goal, after sampling the particles density at ground a NKG lateral distribution is deduced to measure the total number of charged paticles $N_e$ (mostly electrons and positrons) in the shower front. The energy $E_p$ is deduced from the constant intensity cuts method (CIC). The procedure gives a relative error equal to $30\%$. The electric field $E_{0}$ is determined by the lateral distribution function (LDF). The Allan formula has been used to fit the radio LDF with an exponential law (Fig.~\ref{LDFstudies}) following this formula:
\begin{displaymath}
E_i = E_0.exp(-\frac{\sqrt{( (x_i - x_c)^2 + (y_i - y_c)^2 - ((x_i - x_c).cos(\phi).sin(\theta) + (y_i - y_c).sin(\phi).sin(\theta))^2)}}{d_0})
\end{displaymath}
where, $\theta$ and $\phi$ are respectivly zenithal and azimuthal angles reconstructed by a planar fit. This fit has four free parameters $E_{0}$, the LDF decay distance $d_{0}$ and the radio-shower core coordinates on the ground $(x_c,y_c)$, providing the antennas coordinates $(x_i,y_i)$ and the radio filtered pulse amplitude $E_i$ for each antennas. The error on $E_0$ has been estimated through a Monte Carlo analysis. It consist of repeating the LDF fit with $E_i$ values, randomly selected from gaussian probability density function centred on the measured values $E_i$ and a gaussian standard deviation $\sigma_{i}$ took as the RMS of measured radio noise on the antenna. This Monte Carlo method enables to explore the entire phase space and thus to estimate the $E_0$ error on a by event basis. At the end of the procedure a statistical error less than $22~\%$ is deduced. A linear regression is then used to deduce the correlation coefficients in the $(E_p, E_0)$ plan, assuming some hypothesis: (i) The two variables are represented on a linear scale. (ii) Gaussian errors are used on both observables. (iii) The two observables $E_p$ and $E_0$ are assumed to be independent, to cancel the nonlinear covariance term. Results of this procedure is presented on Fig.~\ref{Correlationstudies}. 
It shows that the electric field created at the shower axis $E_{0}$ is clearly correlated with the primary particle energy $E_p$ following the relationship $E_0=a.E_p+b$. This correlation shows that the radio signal emitted by EAS seems coherent. An energy calibration relationship can be deduced, with the following form $E_R=(1/a).E_0-(b/a)$ which allows to have a radio estimation $E_R$ of the energy through the radio method once we have measured $E_0$. It is conceivable that for a future autonomous antennas array, we can measure the primary particle energy only with a pure radio observable.
\section{Radius of curvature reconstruction with CODALEMA}
\begin{figure}[!h]
 \centering
\includegraphics[height=0.3\textheight,width=0.45\textwidth,clip]{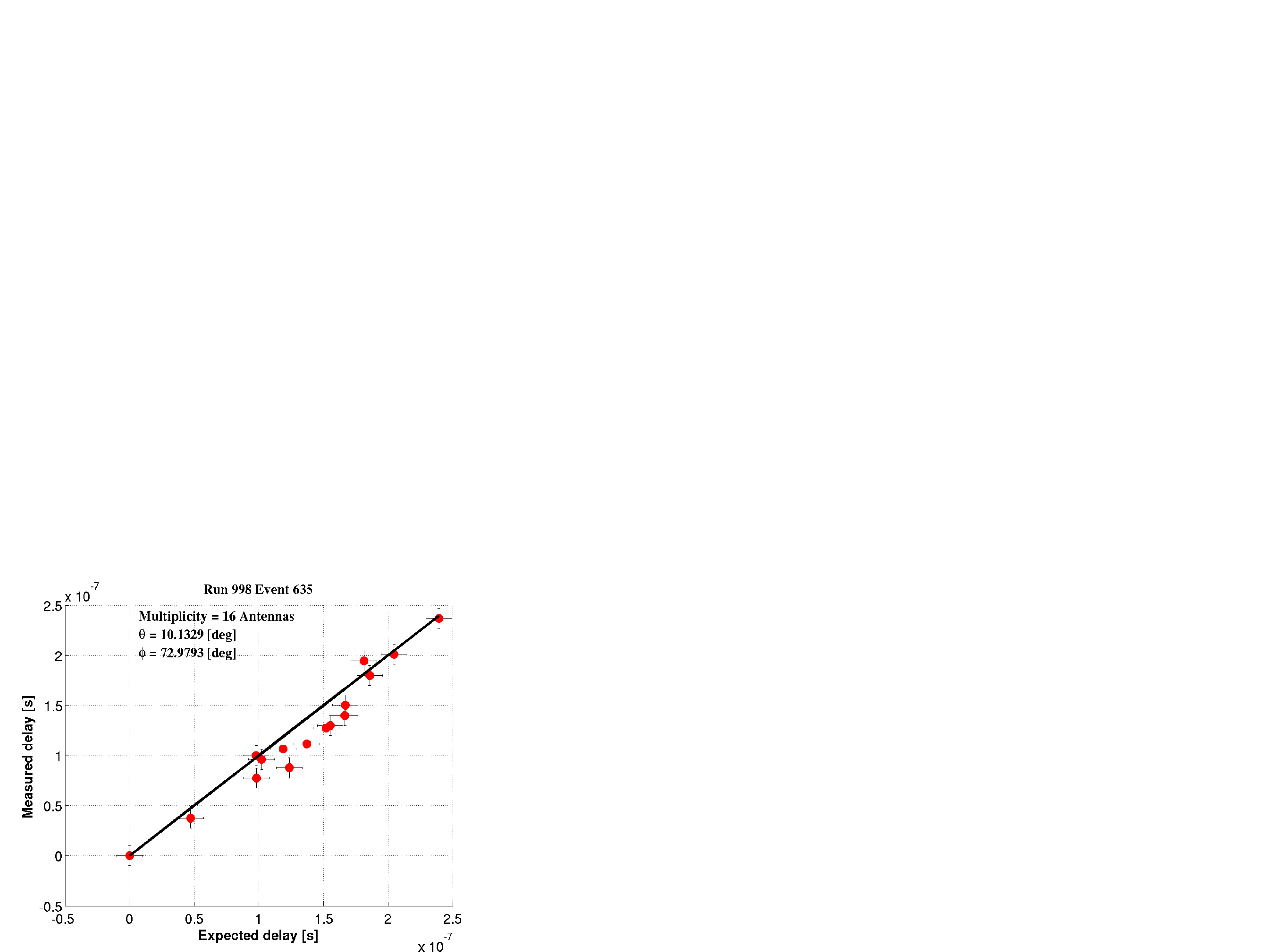}
\includegraphics[height=0.3\textheight,width=0.45\textwidth,clip]{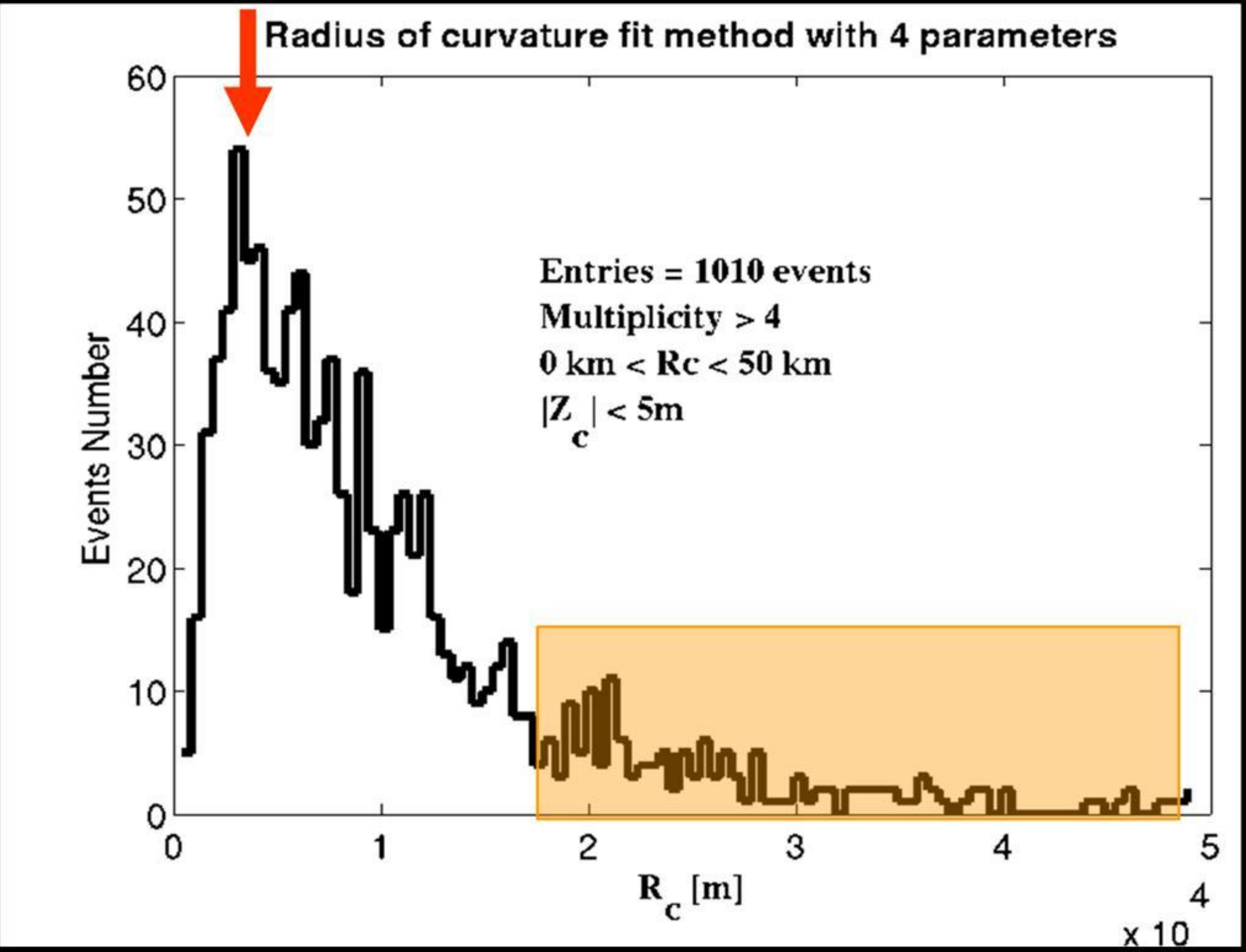}
\caption{{\bf Left:} Expected delay versus measured delay. The black line presents the plane wave fit, Despite the $10~ns$ error bars on both axes, many points are located far from the line, which shows that the wavefront differs from a plane. {\bf Right:} The histogram of the radius of curvature $R_c$ distribution for 1010 selected events. The distribution maximum is located at $4~km$.}
\label{RadiusCurvature}
\end{figure}

%%===
Because theoretical developments indicate that the radio signal shape depends on the shower longitudinal development, it is waited that the wavefront shape provides information on the nature of the primary particle. In the first step of the CODALEMA's analysis, the wavefront was assimilated to a plan determined by a simple planar fit using the arrival times and locations of each tagged antenna. More detailed studies indicates today (Fig.~\ref{RadiusCurvature}. Left) that the measured wavefront differs slightly from the plan in most cases and that it exhibits a curved geometry, favoring the idea of a privileged center for the radio emission during the shower development \citep{RebaiJJC2010,Schroder}. To take into account of these experimental observations, one of the simplest hypothesis is to assume that the maximum of the filtered pulse is linked to a radio signal emission center located along the shower axis. This leads to define a curvature radius $R_c$. Several modelisations suggest that this observable could then be related to the shower maximum, $X_{max}$, which is directly correlated to the UHECR chemical composition. This possibility has been investigated using the present data.
Our fitting method is based on the fit of the residue between the real wavefront determined by arrivals times distribution and the planar wavefront. A parabolic dependence is used to account for the difference. Results of the calculations are presented in Fig.~\ref{RadiusCurvature}. The distribution of the $R_c$ presents a maximum at $4~km$ in global aggrement with the waited characteristic altitude of the emission maximum. However, the tail of the distribution extends up to $20~km$. A present time, the physical interpretation of this long tail is not well understood. This may be due both to a poor estimate of arrival time and to a biased assessment of the estimated error on its time measurement, an arrival time error of $10~ns$ is used. Improvement of these points are underway. 
\section{Conclusions}
Codalema is installed in a radio quiet environment. This advantage has enabled very high accuracy radio signal measurements and several progress in the understanding of the radio emission mechanisms like the geomagnetic field effect or the energy calibration method. A method for the wavefront radius of curvature reconstruction has been presented. The first analysis of the radius of curvature presented shows extremely interesting perspectives with the aim of determining the cosmic rays nature. The deduced location of the emission point are in a satisfactory agreement (for radii less than $10~km$) with the waited values, but for the larger curvatures the values remain poorly understood. Unfortunately the low timing accuracy ($<10~ns$) may limit the analysis with current data. The current array extension which uses standalone radio stations to reach a larger array ($1.5~km^{2}$) will increased the available statistic and improve our interpretations.
\begin{acknowledgements}
This work has been made a part under a grant from R\' egion Pays de la Loire. The author wishes to thank the CNRS (Centre national de la recherche scientifique) for funding his work. The author likes to thank the SF2A week organizers for their hospitality.
\end{acknowledgements}
\bibliographystyle{aa}
\bibliography{rebai}
\end{document}